\documentclass{sig-alternate}

\pdfoutput=1 

\usepackage{cite}
\usepackage{graphicx}
\usepackage{subfigure}
\usepackage{url}
\usepackage{amsmath,amssymb}
\usepackage{bm}
\usepackage[pdftex]{hyperref}

\hypersetup{
	pdfauthor={Fabio Ricciato, Martin Burkhart},
	pdftitle={Reduce to the Max: A Simple Approach for Massive-Scale Privacy-Preserving Collaborative Network Measurements (Extended Version)}
}

\begin{document}
\title{Reduce to the Max: A Simple Approach for Massive-Scale Privacy-Preserving Collaborative Network Measurements\titlenote{This is an extended version of the paper presented  at the \emph{Third International Workshop on Traffic Monitoring and Analysis} (TMA'11), Vienna, 27 April 2011.}}
\subtitle{(Extended Version)}

\numberofauthors{2}
\author{
  \alignauthor Fabio Ricciato \\
  \affaddr{University of Salento, Italy}   \\
  \affaddr{FTW, Austria}
  \alignauthor Martin Burkhart \\
  \affaddr{ETH Zurich, Switzerland}   \\
}

\maketitle              

\begin{abstract}
Privacy-preserving techniques for distributed computation have been proposed recently as a promising framework in collaborative  inter-domain network monitoring.
Several different approaches exist to solve such class of problems, e.g., Homomorphic Encryption (HE) and Secure Multiparty Computation (SMC) based on Shamir's Secret Sharing algorithm (SSS). 
Such techniques are complete from a computation-theoretic perspective: given a set of private inputs, it is possible to perform arbitrary computation tasks without revealing any of the intermediate results. In fact, HE and SSS can operate also on {\em secret inputs} and/or provide {\em secret outputs}. 
However, they are computationally expensive and do not scale well in the number of players and/or in the rate of computation tasks.
In this paper we advocate the use of ``elementary" (as opposite to ``complete") Secure Multiparty Computation (E-SMC) procedures for traffic monitoring.
 E-SMC supports only simple computations with {\em private input and public output}, i.e., it can not handle secret input nor secret (intermediate) output. 
 Such a simplification brings a dramatic reduction in complexity and enables 
 massive-scale implementation with acceptable delay and overhead. Notwithstanding its simplicity, we claim that an E-SMC scheme 
 is sufficient to perform a great variety of computation tasks of practical relevance to collaborative network monitoring, including, e.g., anonymous publishing and set operations. This is achieved by combining a E-SMC scheme with data structures like Bloom Filters and bitmap strings.  
\end{abstract}

\newcommand{\ed}{\color{blue} xxx...}
\newcommand{\edd}{\color{red}}
\newcommand{\ata}{GCR }
\newcommand{\eqdef}{ \stackrel{\textrm{\tiny def}}{=}}

\section{Introduction}
Privacy-preserving techniques for distributed computation have been proposed recently as a promising tool in collaborative  inter-domain network monitoring --- see, e.g., the motivating paper by Roughan and Zhang \cite{RoughanCCR}. In the reference scenario, a set of ISPs are unwilling to share local traffic data due to business sensitivity and/or concerns about their users' privacy. On the other hand, they have a collective interest to perform some global computation on such data and share the final result. 
For example, they might want to aggregate local traffic measurements in order to reconstruct global statistics, and these might be further processed in order to unveal global threats (e.g., botnets) or discover macroscopic anomalies. 
As pointed out already in \cite{sepia10}, each ISP would benefit from comparing its own local view (of traffic conditions) with the global view aggregated over all other ISPs, especially in the occasion of anomalies and alarms, in order to hint at whether the (unknown) root cause is local or global--- a major discriminator for deciding about the reaction. Also, ISPs might be ready to share with other ISPs information about security incidents observed locally,  provided that they can do so anonymously.

Two possible approaches to solve such class of problems are Homomorphic Encryption (HE) and Secure Multiparty Computation (SMC) based on Shamir's Secret Sharing algorithm (SSS for short). 
Both these techniques are ``complete" from a computation-theoretic perspective\footnote{A fully homomorphic, computationally complete HE scheme has been introduced  recently by Gentry \cite{gentry2009fully}. The completeness of SSS is shown in~\cite{benor1988ctn}.}: given a set of {\em private inputs}, it is possible, in principle, to compute any arbitrary function, including structured algorithms involving conditional statements, without revealing any of the intermediate results.
In fact, a distinguishing feature of  HE and SSS is that they can operate also on {\em secret inputs} and/or provide {\em secret outputs} (see the graphical representation in Fig. \ref{fig:whale}).
The notions of {\em secret} and {\em private} are distinct: {\em private data} is known in cleartext to at least one player (and usually only to one), while {\em secret data} remains unknown by all players and can not be reconstructed unless a minimum number of players agree to do so. 
On the other hand, such techniques are computationally expensive --- especially HE --- and therefore do not scale well in the rate of computation tasks (queries) and/or in the number of players.

In this paper we advocate the use of ``elementary" (as opposite to ``complete") SMC procedures for collaborative traffic monitoring.
Such techniques --- hereafter referred to as E-SMC for short --- have a fundamental limit: they support only simple computations with {\em private inputs and public output}, i.e., they can not handle secret input nor secret (intermediate) output. 
We show that such a simplification allows for an enormous reduction in computational complexity and overhead, making such techniques amenable to massive-scale implementation. Notwithstanding its simplicity, we claim that E-SMC is sufficient to perform a broad variety of tasks of practical importance in the field of collaborative traffic monitoring.
In fact, queries  can be chained to build more structured computation tasks (ref. Fig. \ref{fig:dolphin})
whenever intermediate results  --- which are necessarily public in E-SMC --- are not regarded as sensitive. 
Moreover, we show that an additive E-SMC scheme can be combined with local transformations on the private data and/or with particular data structures (e.g., Bloom Filters, bitmap strings) 
in order to extend the range of supported operations. 
  
In this work we take a first step towards unfolding the potential of E-SMC for traffic monitoring. We make three main contributions.
First, we present a simple scheme for E-SMC, called GCR, which is based on additive-only or multiplicative-only secret computation and extends an idea presented earlier in \cite{Atallah04}.
Second, we highlight some system-design aspects of \ata that enable massive-scale implementation: in particular, we propose to split the computation into 
 {\em offline} randomization and {\em online} aggregation phases.
Third, we describe how  GCR can support a number of operations relevant to collaborative traffic monitoring --- like set operations, anonymous publishing and anonymous scheduling --- when combined with data structures like Bloom Filters and bitmap strings. 

The aim of this report is not to provide definitive results nor quantitative assessments, but rather to indicate a direction of work to researchers engaged in inter-domain traffic monitoring.
We claim that a broad variety of tasks  of practical relevance to this field do not necessitate to resort to  ``complete" (and complex) privacy-preserving schemes but can be satisfactorily attained by E-SMC. Thanks to their simplicity, collaborative systems based on E-SMC are amenable to massive-scale implementation, with very large number of players and/or very high rate of queries. In turn, system scalability paves the way towards  customer-driven collaborative monitoring, where participating players do not map to ISPs but rather to their customers --- think, e.g., to mid-to-large companies with own IT security staff. 
 This is indeed a new avenue of collaborative network monitoring that might have in E-SMC its enabling technology.


\newcommand{\mipicwidth}{0.47\textwidth}
\begin{figure*}[tb]
\centering
\subfigure[]{ \includegraphics[width= \mipicwidth]{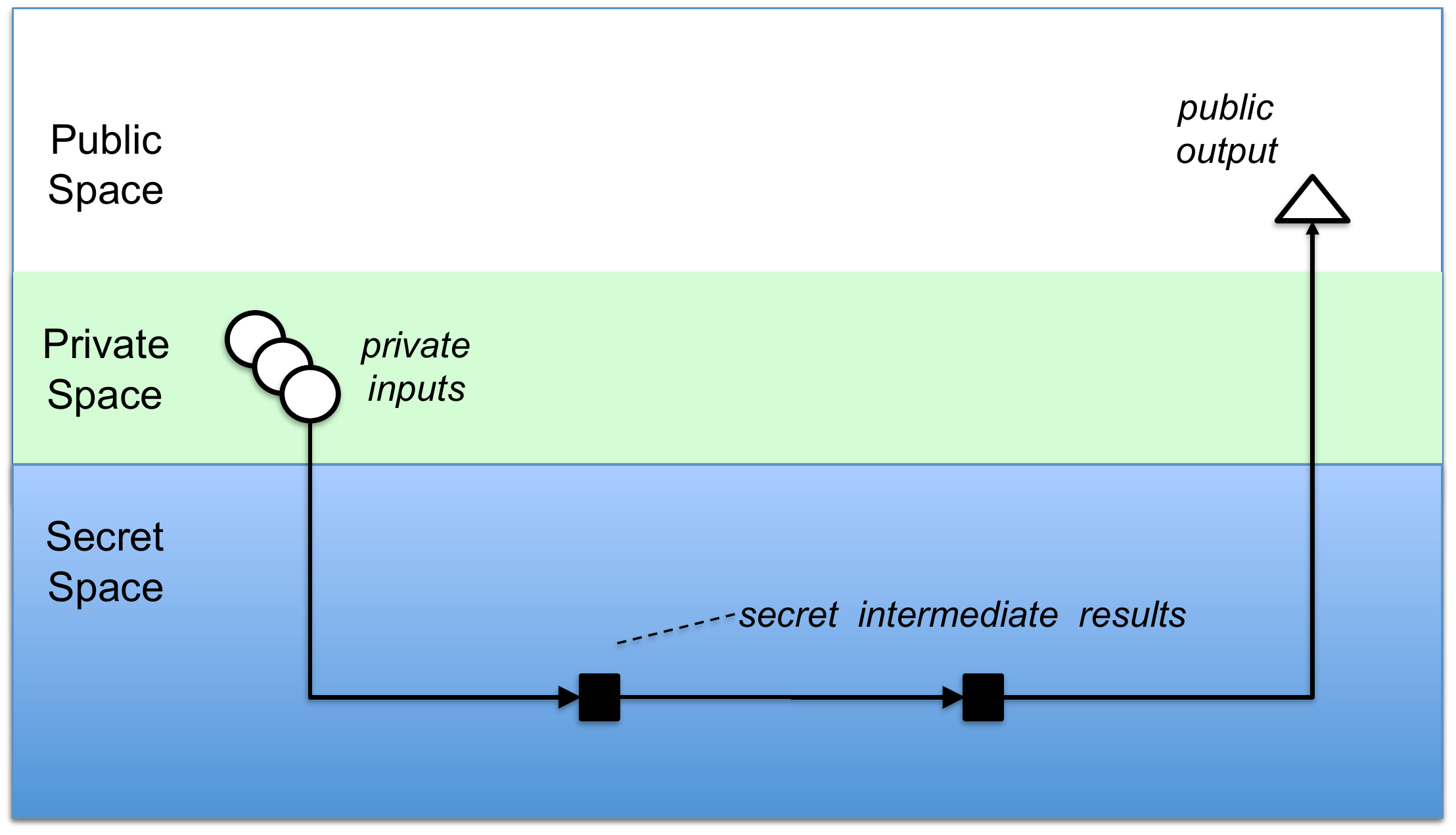} \label{fig:whale} }
\subfigure[]{  \includegraphics[width= \mipicwidth]{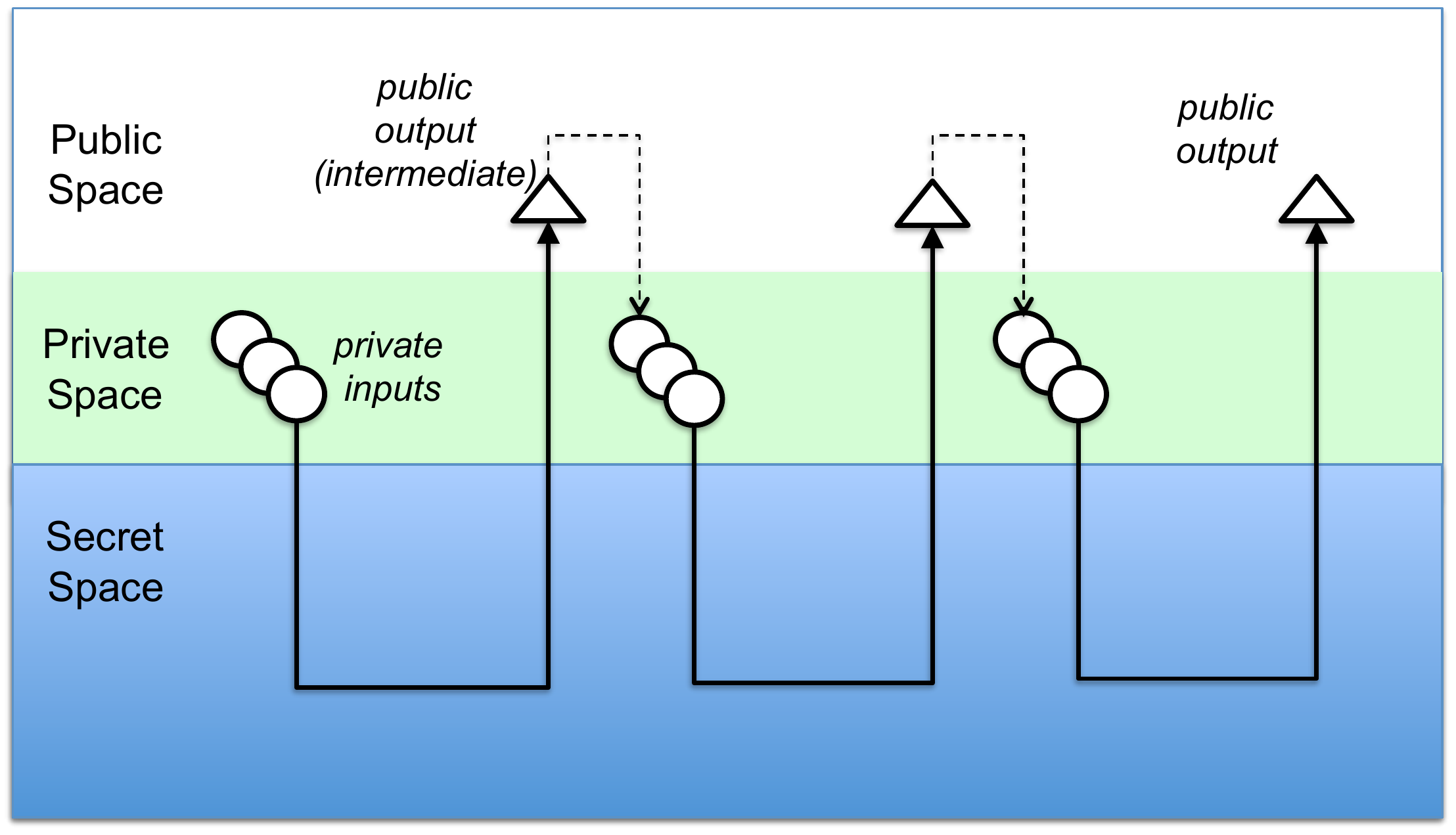}  \label{fig:dolphin} }
\caption{ Graphical representation of a ``complete" secure procedure with secret intermediate results (a) and a sequence of ``elementary" secure operations chained by public intermediate results (b). } 
\label{fig:whale-dolphin}
\end{figure*}

\section{The \ata method}

We consider the classical SMC scenario where a set of $n$ players collaborate to compute a function of some private data --- e.g., traffic statistics,  network logs, records of  security incidents.   
As customary in SMC, we assume a \emph{semi-honest} model (also known as \emph{honest-but-curious}):  {all} players cooperate honestly to compute the final result, but a subset of them might collude to infer private information of other players. 
In other words, no {\em malicious} player will attempt to interrupt nor corrupt the computation process, e.g., by providing incorrect input data.

In this section we present a simple method to perform secure private {\em addition} which extends an idea presented earlier by Atallah {et al.} in \cite[\S4.1]{Atallah04} based on additive secret sharing.
We refer to our method as ``Globally-Constrained Randomization", \ata for short. 
We show that
 GCR, which is simple conceptually, lends itself very well to massive-scale implementation.
We propose also for the first time a variation of the scheme to perform secure {\em multiplication}.

\subsection{Notation}\label{sec:notation}
We consider a set of $n$ players $\{P_i, \, i=1 \ldots n\}$ with $n \geq 3$ (normally $n>>1$). 
The maximum number of colluding players will be denoted by $l$ (collusion threshold) with $l \leq n-2$.
Note that $l$ is a design parameter that can be set independently from the system size $n$.
For each computation task (query)  each player $P_i$ involves two elements:
\begin{itemize}
\item $a_i$ is the 
 \emph{private input} of  $P_i$ to the summation. For some queries, it is obtained by applying a local transformation $g()$ on some other inner private data $b_i$, i.e., $a_i = g(b_i)$.
\item $r_i$ is the private {\em random element} which $P_i$ has previously generated cooperatively with other players in the way presented later.
\item $v_i \eqdef a_i + r_i$ is the {\em public input} which $P_i$ eventually announces to the other players.
\end{itemize}
The collection of random elements across all players constitutes a  Random Set (RS) and will be denoted by ${\bf{r}} \eqdef \{r_i, \, i=1 \ldots n \}$.
The goal of the computation round is to obtain the {\em public output} result $A \eqdef f(a_1,a_2 \ldots a_n) = f(g(b_1),g(b_2),..g(b_n))$ 
without disclosing the values of the individual $a_i$'s. 
For each computation, all input elements ($a_i, r_i,v_i$) and the output $A$ must be in the same format.
For the additive scheme they  must be defined over the same 
{\em additive commutative group} (Abelian group).
We will consider the following distinct cases:

\begin{description}
	\item [Real scalars:]  
	$a_i, r_i$ and $A$ are real numbers defined in the interval $\mathbb{R}_p \eqdef [0,p]$. For the sake of simplicity we will assume $p$ integer, but not necessarily prime. 
The group operation in this case is modulo-$p$ addition.
A generic random element $x$ is a random value extracted uniformly in $[0,p]$, i.e., $x \sim \mathcal{U}(0,p)$. The null element is the zero value.
	\item[Integer scalars:] 
this is a sub-case of the previous one, where $a_i, r_i$ and $A$ are integers in $\mathbb{Z}_p \eqdef [0,p]$. Unless differently specified, $p$ is  not necessarily a prime number. In practice, it is convenient to choose $p=2^q$ ($q$ integer) so that modulo-$p$ addition maps to wrap-around of a $q$-bit counter. 
	\item[Binary strings:] $a_i, r_i$ and $A$ are binary strings of length $k$. 
The group operation is therefore bitwise addition (XORing).
In this context a generic random element  $x$ is a random string, i.e., a collection of bits set randomly to 1 or 0 independently and with equal probabilities. The null element is  a string with all '0's.
	\item[Arrays of counters:] $a_i, r_i$ and $A$ are vectors of $k$ elements, and each element is a  $q-$bit counter.
The group operation is therefore an array of  $k$  parallel modulo$-p$ additions.
In this context a generic random element  $x$ is a collection of $k$ random values $<x_1, x_2,..x_k>$ extracted independently and uniformly in $[0,p-1]$. 
The null element is an array of zeros. 

\end{description}

The format of the input elements $a_i, r_i$, the exact values of the parameters (e.g., $k,q$) and, if applicable, the choice of the transformation function $g()$ depend on the particular kind of operation (query) as detailed in \S\ref{sec:operations-advanced}. 
 In the following we will use the symbol  `$+$' to refer generically to the addition between two terms and  `$\sum$' for multiple terms, without specifying the group operation.

\subsection{Description}\label{sec:ata}

The central aspect of the \ata method is that RS is constructed in a way that guarantees the zero-sum condition, i.e., the composition of random elements across {\em all} users sums up to the null element: 
\begin{equation}
\sum_{i=1}^{n}{r_i} = \bf{0}. \label{eq:zero-forcing}
\end{equation}
Moreover, the generation of RS ensures that the individual $r_i$'s can not be inferred by other players --- provided that the number of colluding players remains below the colluding threshold $l$. 
Each player $P_i$ then shares with other players (e.g., via a central collector) the sum of data plus random elements, i.e., $v_i=a_i+r_i$, which serves as the public input to the computation.
When {\em all} input elements $v_i$ are collected, the value of $A$ is obtained by summing them all, formally:
\begin{equation}
\sum_{i=1}^{n}{v_i} = \sum_{i=1}^{n}{(a_i + r_i)} = \sum_{i=1}^{n}{a_i} + \sum_{i=1}^{n}{r_i} = A + 0 = A
\end{equation}
Note that the value of $A$ can be reconstructed only when the inputs from {\em all} players have been collected: it is sufficient that a single player (among those that have contributed to generate the RS ${\bf{r}}$) fails to provide its input element to prevent the computation of $A$. This is the main disadvantage of \ata compared to SSS, as discussed later in \S \ref{sec:comparison-ata-sss}.\\

{\bf RS generation}
Hereafter we describe how each generic player $P_i$ ($i=1 \ldots n$) constructs its random element $r_i$ in cooperation with other players, so as to collectively build the RS  $\bf{r}$.
Note that the RS generation procedure can be run {\em in parallel} by all players and is completely asynchronous.
Each random element is initially set to the null element, i.e., $r_i=0$. 
Each player $P_i$ extracts $l+1$ random variables $x_{i,j}$ ($j=1 \ldots l+1$) and computes their sum $y_i \eqdef \sum_j{x_{i,j}}$.
It calculates  the additive inverse\footnote{In modular arithmetic  the additive inverse $\overline{y}$ of $y$ is the element that satisfies $\overline{y}+y = 0$. For real numbers in $[0,p]$, $\overline{y}= p-y+1$, while for binary strings $\overline{y}=y$. } $\overline{y_i}$ of $y_i$ and adds it to its own random element, i.e., $r_i \leftarrow r_i + \overline{y_i}$.
At the same time, $P_i$  contacts $l+1$ randomly selected other players and  sends one variable $x_{i,j}$ to each of them: 
each contacted player $P_j$ will then increment its random element by  $x_{i,j}$, i.e., $r_j \leftarrow r_j + x_{i,j}$.
This method is secure against collusion of up to $l$ players.
Notably, the value of $l$ is a free parameter,   independent from the system size $n$, which can be tuned to trade-off communication overhead with robustness to collusion --- both scale linearly in $l$.\\

{\bf Computation phase}.   
With \ata the computation is basically a summation over $n$ public inputs, the $v_i$'s,  and 
no particular constraint applies to the aggregation method which can be centralized or distributed.
For the sake of simplicity,  we assume in the following a fully centralized scheme, with a single master --- not necessarily a player ---  that is in charge of launching the query, collecting the $n$ public inputs, computing the result and finally publishing it to all the players.  
Another possible option is tree-based aggregation:  players are arranged into a tree, where each node collects the inputs from its children and sends the summation result to its parent node, until the root computes and publishes the final result. 
More sophisticated peer-to-peer method can also be adopted at the cost of some additional coordination overhead. 
The point to be taken is that the \ata method is oblivious to the particular input aggregation scheme. 

\subsection{Extension to multiplication} \label{sec:multiplication}
It is straightforward to adapt the \ata scheme to support multiplication of positive integers.
First, the input and output data $a_i, r_i$ and  $A$  must be defined over the multiplicative group  $[1,p]$ with $p$ a prime number:
primality guarantees that each element has a unique {\em multiplicative} inverse element 
(note the difference with {\em additive} \ata which does not require primality of $p$).
Second, all modulo-$p$ additions are replaced by modulo-$p$ multiplications.
Third, the balancing constraint eq.  (\ref{eq:zero-forcing}) is replaced by:
\begin{equation}
\prod_{i=1}^{n}{r_i} = 1 \label{eq:one-forcing}
\end{equation}
In this way we obtain a multiplicative variant of the additive sharing scheme, which to the best of our knowledge was never considered in previous literature. 
It is important to remark that \ata can support either addition {\em or} multiplication, but it can not compose addition {\em and} multiplication operations without reconstructing and resharing values. In the secret evaluation scheme, it is therefore not computationally complete.

Finally, note that multiplicative GCR can not take zero as private input, as that would automatically force to zero also the public output, i.e., $a_i=0 \Rightarrow v_i=a_i \cdot r_i=0 \; \forall r_i$, therefore leaking the private value. 
In practice,  before launching a secret multiplication, one can easily check for the presence of zero inputs, e.g., with a preliminary  round of Conditional Counting (see \S \ref{sec:counting}).

\subsection{Sensitivity of Output}
It is important to note that SMC in general (not only E-SMC) only guarantees that no information is leaked \emph{from the computation process}. 
That is, it solves the problem of \emph{how} to compute a function $f()$ on distributed data in a privacy-preserving way. 
An orthogonal problem is to find out \emph{what} is safe to compute.
Just learning the resulting value $f()$ could allow the inference of sensitive information.
For example, if the private input bits must remain secret, computing the logical \texttt{AND} of all
input bits is insecure in itself: if the final result was $1$, all input bits must be $1$ as well and are thus no longer secret.
In SMC, \emph{it is the responsibility of the input providers to verify that learning $f()$ is acceptable}, 
in the same way as they have to verify this when using a trusted third party.
While with SMC, this analysis has to be performed for the final result only, in E-SMC it has to be
performed individually for each step computing public intermediate results.

A recently suggested approach to deal with this is \emph{differential privacy}~\cite{dwork2008differential,differentialPrivacy2010}, 
which systematically randomizes answers to database queries to prevent inference of sensitive input data.
If data records are independent, it guarantees that it is statistically impossible to infer the presence or absence of 
single records in the database from answers to queries.
Differential privacy and SMC complement each other very well. Using differential privacy, it is possible to 
specify a randomized output $\widetilde{f}()$ that is safe for public release. Using SMC,
it is possible to actually compute $\widetilde{f}()$ in a privacy-preserving manner, without relying on a trusted third party.
Intuitively, the stronger $f()$ aggregates input data, the less randomness needs to be added.

\section{System-design considerations}\label{sec:design}

In this section we consider a number of system-level aspects. In particular, 
we propose to split the \ata operation into an \emph{offline} generation of RS and \emph{online} aggregation phase, and show how joins and leaves of nodes can be  handled efficiently.
We also compare the \ata scheme to Shamir's secret sharing scheme, which, among the existing alternatives for performing SMC, allows the most efficient solutions.

\subsection{Offline generation of Random Sets} 
One key advantage of \ata is that the process of generating the RS is completely decoupled --- and can be run independently --- from the actual computation round.
 This  has important implications for the design of a massive-scale system, enabling efficient management of  the communication load and minimal response delay. 
We devise  a system where lists of RS are generated {\em offline} and stored for later use. 
At any time, each player $P_i$ has available a collection of random elements $r_i[u]$, indexed in $u$, which can be readily used for future computation rounds. 
The communication protocol must ensure that the RS indexing is univocal and synchronized across all players. 
During the {\em online} computation phase, the query command broadcasted by the central master will indicate explicitly the RS index to be used for the production of the public inputs $v_i$'s. 

Performing  RS generation offline brings several advantages. First, it minimizes the query response delay down to the same value of an equivalent cleartext summation.
Second, it allows to reduce the impact of communication overhead onto the network load by scheduling the RS generation process in periods of low network load (e.g., at night or week-end). Moreover, generation of multiple RS can be {\em batched}, meaning that in a single secure connection (typically SSL over TCP) two players can exchange multiple $<$variable,index$>$ pairs  $\{x_{i,j}[u], u \}$ which collectively build a collection of RS $\{\mathbf{r}[u]\}$. This greatly reduces the communication overhead associated to connection establishment (handshaking, authentication, key exchange, etc.).\\

\subsection{Joining and leaving}
In the \ata scheme, the set of players participating in the computation round must match exactly the set of players that have previously built the RS: the final result will not be reconstructed if the two sets differ by even a single element.
If RSs are generated offline,  the set of players might have changed during the interval between the generation of  ${\bf{r}}[u]$ and its consumption in a query.
It would be very impractical to trash all pre-computed RSs upon every new player joining or leaving the system --- an event not infrequent for systems with many players.
Fortunately this is not necessary and each legacy RS can be incrementally adjusted upon new join or leave with only $l+1$ operations.

When a new player $P_i$  joins the system, it learns from other players the index range currently in use $ \{u_1 \ldots u_2\}$ (note this information is public) and computes a set of random variables $x_{i,j}[u]$ for $j=1 \ldots l+1$ and $u \in \{u_1 \ldots u_2\}$. It then sets its local random elements as $r_i[u] = \overline{y_i}[u]$ (recall that $y_i = \sum_{j=1}^{l+1}{x_{i,j}[u]}$). Then for each index value $k$ it selects $l+1$ other players to which it sends the individual variables $x_{i,j}[u]$. 
Similarly, when an existing player $P_i$  wants to leave the system, it must first ``release" its random elements  $r_i[u]$. 
The simplest way to accomplish that is to simply pass the value of $r_i[u]$ to another randomly selected player $P_j$ and let the latter update its local random element as $r_j[u] \leftarrow r_j[u] + r_i[u]$. 
Note that we are assuming a ``cooperative leaving" behavior: players release their unused random elements to the system before leaving. 
However if a player shuts down without releasing its random elements --- e.g., due to failure, power off or disconnection ---  all RSs in the entire system are invalidated and become useless. In large scale systems such events might not be infrequent, and proper countermeasures must be adopted to minimize their impact (e.g., node redundancy). 

\subsection{\ata versus Shamir's Scheme}\label{sec:comparison-ata-sss}

We now compare \ata to Shamir's secret sharing scheme~\cite{shamir79}, denoted by SSS.
E-SMC, along with all the use cases described in the following sections 
can be implemented with either \ata or SSS. 
In GCR, reconstruction of public values is implicitly done after each processing step, while in SSS reconstruction needs to be scheduled explicitly if desired.

In SSS, a secret value $s$ is shared among a set of $n$ players by generating a random
polynomial~$f$ of degree~$t<n$ over a prime field $\mathbb{Z}_p$, such that $f(0)=s$.  Each player $i=1\ldots n$ then receives an evaluation
point $s_i=f(i)$, called the share of player $i$. The secret $s$ can be reconstructed from any~$t+1$ shares using Lagrange interpolation but is completely undefined for
$t$ or less shares. 
Because SSS is linear, addition of two shared secrets can be computed by having each player locally add his shares of the two values.
Multiplication of two shared secrets requires an extra round of communication to guarantee randomness and to correct the degree of the
new polynomial~\cite{gennaro1998sva}. Thus, a distributed multiplication requires a synchronization round with $n^2$ total messages. For multiplications to work, the
degree must be such that $n \geq 2t+1$.

There are two main advantages of SSS over \ata. First, the basic operations for addition and multiplication accept public, private, and 
also \emph{secret} input data and output \emph{secret} data. That is, even without reconstructing intermediate values, it is possible to arbitrarily compose secret operations, corresponding to Fig.~\ref{fig:whale}. The \ata scheme allows composition of addition and multiplication only if intermediate results are publicly reconstructed, because the sharing operation to be applied (additive or multiplicative) depends on the next operation type. The second advantage of SSS is that it realizes a $(t+1)$-out-of-$n$ threshold sharing scheme. That is, any set of $t+1$ players can reconstruct a secret, being robust against up to $n-t-1$ ``missing'' players. In \ata, a single non-responsive player renders reconstruction of secret information impossible. 

While E-SMC can also be implemented with SSS, \ata is highly optimized for \emph{online} processing of queries. SSS
requires linear storage overhead ($n$ shares to be stored for each secret value), whereas \ata has constant storage overhead (one random value per private input). When processing the query, \ata involves zero communication overhead, since the players just send their randomized values instead of the original value to the aggregation node(s). 
In SSS, when $n$ players want to sum up their values, each of them generates $n$ shares ad-hoc and distributes them to the others. 
In principle, the players could pre-generate $t$ random shares and distribute them in a pre-processing phase. In the online phase, they would calculate the remaining $n-t$ shares using Lagrange interpolation, such that the interpolated polynomials represent their actual secrets. However, after distributing the last shares, each player still needs to perform $n-1$ additions locally and for final reconstruction, send their shares of the sum to the aggregation node(s), which eventually interpolates the final polynomial. 
It is not obvious how to further split this process into a offline pre-processing and an online phase similar to GCR, where a single message and addition operation is enough.

Another advantage of \ata is that the additive scheme is not restricted to prime fields. This allows to set the field size to $2^{32}$ or $2^{64}$ and therefore use implicit 32 (64) bit register wrap-arounds of CPU operations instead of performing an explicit modulo operation\footnote{In general, $mod(a, n) = a - n * floor(a / n)$, which uses an additional division, multiplication, and subtraction operation.}. 
Furthermore, the multiplicative \ata scheme does not need an additional synchronization round like SSS.  

In summary, provided that intermediate results are not sensitive, \ata allows for a much smaller storage and computation overhead during the online processing phase.

\section{Basic Operations}\label{sec:operations}
Here we briefly sketch some basic operations that can be mapped to a secure addition
with a public parameter and/or a public conditional statement.
As such, they can be accomplished directly by   \ata method  or any other scheme for secure addition.

\subsection{Summation}\label{sec:summation}
The summation of positive real scalars $A=\sum_i{a_i}$, with $a_i \in [0,p]$, is performed directly as explained above via modulo-$p$ additions. The only significant  constraint is on the value of $p$ which must be greater than the total sum, i.e., $p > A$. 
The method can be easily extended to handle negative elements defined in $[d_1,d_2]$, with $d_1 < 0 < d_2$, by imposing a fixed shift $+|d_1|$ to all inputs $a_i$'s and then subtracting $n|d_1|$ from the output.  
 Note however that summation of negative numbers is unusual in  traffic monitoring.

\subsection{Conditional Counting}\label{sec:counting}
We consider two versions of Conditional Counting (CC) queries: ``player counting"  and ``item counting".
In the first version, the goal is to count how many players match a public condition $\mathcal{C}$ which is
explicitly announced as a public query argument.
Each player $P_i$ sets $a_i$ to 0 or 1 depending on whether or not it matches the condition $\mathcal{C}$.
Therefore CC maps to a particular case of summation, where $a_i \in \{0,1\}$ and $p  \geq n+1$.  
In the ``item counting" version instead the goal is to count the total number of items (e.g., hosts or alarm records) matching the condition $\mathcal{C}$, where multiple items might be observed by a single player. Again, counting maps directly to summation of integers.

CC queries can serve as a preliminary round to other more advanced queries, e.g., to identify the presence of zero inputs before multiplication (see \S \ref{sec:multiplication}), or to discover the exact number of active players  before a round of Anonymous Scheduling (see \S \ref{sec:scheduling}). 

 \subsection{Histograms and max/min discovery}\label{sec:binhist}
Each player $P_i$ has a scalar private value  $b_i$ 
and the problem is to derive a $K$-bins histogram of the distribution of the $b_i$'s. This can be easily achieved 
   by using  CC queries, indexed in $k$, with condition $\mathcal{C} := Y_{k-1} < b_i \leq Y_k$, wherein the threshold values $\{ Y_k, k=1 \ldots K\}$ represent the bin boundaries. 
The number of CC queries is equal to the number of bins $K$. However since bin boundaries are pre-determined, the queries can be batched in a single round using an array of $K$ counters. 

In a similar way it is possible to discover the maximum value of the $b_i$'s.
Again, one can resort to a sequence of  CC queries where the threshold values $Y_k$ are adjusted dynamically based on the previous result following a binary search.  
If $b_i$'s are integer and upper bounded by $p$, the maximum is found in $\log_2{p}$ rounds. 
Note however that the results of all intermediate queries are public, therefore this method discloses more information about the $b_i$'s distribution than just the maximum. 
In a similar way it is possible to discover the minimum. 

\section{Advanced Operations}\label{sec:operations-advanced}

Here we show a few examples of more advanced operations which can be mapped to E-SMC queries in combination with specific constraints on the input data elements and/or a proper local transformation function $g()$. 
For each of them we illustrate a possible application for  collaborative network monitoring.
This section is one of the main contributions of the paper: to the best of our knowledge we are the first to ``interpret" the following operations as applications of SMC using the additive sharing scheme.

\subsection{Multiplication}
Multiplication of positive integers can be accomplished directly by the multiplicative version of \ata presented in \S \ref{sec:multiplication}.
Alternatively, 
the multiplication of positive real numbers $B=\prod_i{b_i}$ (for $b_i > 0$)  can be mapped to a summation in  the logarithmic domain. Each player locally computes $a_i = \log_c{b_i}$ 
 and then the computation proceeds as a simple summation of real numbers, leading to $A = \sum_i{a_i}$. Finally, the result is computed as $B=c^{A}$. 
 Some numerical issues might arise when the product involves a large number of non-unitary terms, due to the accumulation of rounding errors in the representation of the logarithmic values --- these however are well studied problems.

\subsection{Set Operations}\label{sec:setop}

In this section, we first describe how (probabilistic) set operations can be implemented using bloom filters with any SMC scheme that supports both, private additions and multiplications (e.g., SSS). 
We then outline what subpart of that functionality can easily be implemented with GCR. 

Bloom filters (BF) are powerful data structures for representing sets~\cite{broder2004network}. 
A bloom filter for representing a set $S = \{x_1, x_2, \ldots, x_n\}$ of $n$ elements is described by an array
of $m$ bits, initially all set to $0$. The BF uses $k$ independent hash functions
$h_1, \ldots , h_k$ with range ${1, \ldots , m}$. For each element $x \in S$,
the bits $h_i(x)$ are set to 1 for $1\leq i \leq k$.
For checking whether an element $y$ is a member of $S$, we simply check whether all bits $h_i(y)$ are set to $1$.
As long as the BF is not saturated, i.e., $m$ is chosen sufficiently large to represent all elements, the total number of non-zero buckets allows to 
accurately estimate $|S|$.
Counting Bloom Filters (CBF) are a generalization of BFs, which use integer arrays instead of bit arrays. Thus, CBFs allow to represent \emph{multisets}, in which each element can be represented more than once.
Note, that while a (C)BF allows to efficiently check for element membership, it can not be used to enumerate the contained elements, in general.
Compared to state-of-the-art approaches for privacy-preserving set operations, which use homomorphic encryption (e.g.,~\cite{kissner2005pps}), this allows for very efficient and scalable solutions.

\paragraph{Set Union}
If each player $i$ has a local set $S_i$, they can construct the union of their sets $S=S_1 \cup S_2 \cup, \ldots, \cup S_n$ by performing private OR ($\vee$) over their BF arrays. 
If inputs are multisets, represented by CBFs, the aggregation operation is addition instead of OR.
Using CBFs, each player can learn  the number of occurrences of specific elements across all players or the number of other players that report each element (by using a BF as input). From the aggregate CBF, one could, for instance, compute the entropy of the empirical element distribution.

\paragraph{Set Intersection}
In order to perform set intersection on BFs, the players simply use the AND ($\wedge$) operation for aggregating their sets $S=S_1 \cap S_2 \cap, \ldots \cap S_n$.
Only buckets set to $1$ in all the players' BFs will evaluate to $1$ in the aggregate BF. In this specific scenario, it is also possible for each player $i$ to enumerate all elements in $S$ simply by iterating over all $x \in S_i$ and checking whether $x \in S$, since $S \subseteq S_i$.

\paragraph{Set Operations with GCR}
GCR directly supports the addition operation and therefore set union on multisets. If the counts in each bucket are not sensitive, the union and intersection of sets can be computed from the public union of multisets --- the intersection, for instance, is given by selecting all elements with count $n$. 
However, private union and intersection  directly on sets can not be delivered by GCR.
In fact, union requires OR, i.e., a combination of addition and multiplication\footnote{Note that with $a,b$ being bits, $a \vee b = a+b-2ab$ and $a \wedge b = ab$.} not supported by GCR, while the problem with intersection is that  multiplicative GCR does not include $0$ (see \S\ref{sec:multiplication}).

\subsection{Anonymous publishing}\label{sec:publishing}

The goal is to let one player $P_1$ publish to all other players a binary string $w$ without revealing its identity. 
The string $w$ can be, for example, a malware payload that $P_1$ has discovered with an IDS, or the description of an attack which was observed locally.
Moreover, $w$ could be used as a public condition for a future Conditional Counting round  (\S \ref{sec:counting}), e.g., to discover how many other players have observed the same event.
There are several reasons why the publisher wants to remain anonymous.
First, knowing that it was hit by the malware might be detrimental to its reputation among customers.
Second, such information might benefit other potential attackers.

DC-nets~\cite{chaum1988dining} are a basic and unconditionally secure solution for anonymous publishing. In the following, we devise an alternative solution that does not require pair-wise shared secrets, 
 and deals with the problem of i) detecting collisions and ii) scheduling the publication process to avoid collisions. 

Let $k$ denote the length of string $w$, and denote by $C(w)$ a Cyclic Redundancy Check (CRC) control field of length $c$ computed on $w$ --- the need for CRC is explained below.
It is straightforward to map an Anonymous Publishing round to a bit-wise summation on strings of length $k+c$.
The publisher $P_1$ sets its data element to the concatenation of $w$ and $C(w)$, i.e., $a_1=<w,C(w)>$, while all other players
set their data elements to null ($a_j=0, \; j \neq 1$).
Therefore the public result will return the string $w$ in cleartext, i.e., $A=a_1=<w,C(w)>$, but since the individual data elements remain unknown the identity of the publisher can not be reconstructed.
Such a simple approach works only if exactly {\em one} player attempts to publish in the computation round: if two (or more) players $P_1$ and $P_2$ attempt to publish different strings, we have a collision --- i.e., the computed result will be the combination $A=<w_1\oplus w_2, C(w_1) \oplus C(w_2)>$ ('$\oplus$' for bit-wise summation) from which neither of the elements $w_1,w_2$ can be derived. 
However the collision can be easily revealed by CRC failure as in general $C(w_1+w_2) \neq  C(w_1) \oplus C(w_2)$.
The ``collision recovery" procedure can simply foresee the repetition of new anonymous publishing rounds associated to a back-off scheme to avoid that the same players collide again in the next round --- a mechanism conceptually equivalent to Slotted-Aloha.

A simple ``detection and recovery" approach is not effective when   
the instantaneous rate of publishing attempts is high --- this is of particular concern in large-scale system with many players ($n>>1$) and/or in presence of correlated attempts (e.g., a spreading malware payload caught simultaneously by different domains).   
In such cases it is preferable to adopt a ``collision prevention" method by 
orderly scheduling the publishing rounds for different players.
This can be achieved by a single round of anonymous scheduling, as explained below. 

\subsection{Anonymous scheduling}\label{sec:scheduling}
The problem is defined as follows. 
Out of the total $n$ players, a subset of $m<n$ ``active" players are ready to perform a given action, e.g., anonymous publishing. 
The problem is then to schedule the $m$ active players {\em without knowing nor revealing their identities}. 
This apparently difficult task can be easily accomplished by bit-wise summation over strings of size $k>>m$. 
At the query round, the inactive players set their data elements to the null string, while each {\em active} player $P_i$ extracts uniformly a random integer $q_i \sim \mathcal{U}(1,k)$ 
 and then builds its data element $a_i$ with a single '1' at the $q_i$-th position and all other bits set to '0'. 
The bitmap length $k$ must be set large enough to ensure that bit-collision probability --- i.e.,  two or more players independently picking the same random value $q_i$ --- is kept acceptably low. 
  
Assuming that no bit-collision has occurred, the final (public) result $A$ is a bitmap with $m$ '1's and $k-m$ '0's. 
Upon learning $A$, each active player $P_i$ checks whether the bit in the $q_i$ position is set to '1', and if so it counts the number of '1's in the preceding positions, say  $\mu_i$, from which he learns it has been scheduled  
  in the successive $(\mu_i+1)-$th query round. 
If otherwise the $q_i$-th bit is  '0', $P_i$ infers that a collision has occurred and waits for the next scheduling round.

Note that in case of bit-collisions the round does not completely fail: if collisions involves only two (or any even number of) players, the colliding players will simply wait for the next scheduling query. If three (or any odd number of) players have collided on the same $q-$th bit, they would again collide in the $q-$th query round. However this is not a serious problem as far as collisions in the query rounds can be detected and recovered (e.g., by CRC failure in case of Anonymous Publishing). 

The number of active players $m$ is relevant to the setting of the bitmap length $k$ ($k>>m$). One conservative approach is to simply assume the worst case $m=n$. Alternatively, a preliminary Conditional Counting query (\S \ref{sec:counting}) might be launched to discover the exact value of $m$.  The latter approach has also another advantage: with knowledge of $m$, the occurrence of bit-collisions can be easily revealed by comparing to the number of '1's in the final result, i.e., $|A|_1$. In fact, the difference $m-|A|_1$ equals to the number of colliding players. For example, $m-|A|_1=1$ implies that only a two-player collision has occurred, 
 and the master can decide to validate the current scheduling round --- implicitly deferring the two colliding players to a future scheduling round ---  or to invalidate it and immediately re-launch a new scheduling round.

\section{Related Works}

SMC is a cryptographic framework introduced by Yao~\cite{yao1982psc} and later
generalized by Goldreich et al.~\cite{Goldreich1987any}. 
SMC techniques have been widely used in the data mining community. For a comprehensive survey, please
refer to~\cite{Aggarwal2008survey}.
Roughan et al.~\cite{RoughanCCR} first proposed the use of SMC techniques for a number of applications
relating to traffic measurements, including the estimation of
global traffic volume and performance measurements~\cite{roughan2006ppp}.
In addition, the authors identified that SMC techniques can
be combined with commonly-used traffic analysis methods and tools, 
such as time-series algorithms~\cite{Atallah04} and sketch data structures. 

However, for many years, SMC-based solutions have mainly been of theoretical interest due
to impractical resource requirements. Only recently, generic SMC frameworks optimized for
efficient processing of voluminous input data have been developed~\cite{sepia10, Sharemind}.
Today, it is possible to process hundreds of thousands of elements distributed across
dozens of networks within few minutes, for instance to generate distributed top-k reports~\cite{burkhart-topk}.
While these results are compelling, they stick to the completely secret evaluation scheme.
Our work aims at boosting scalability even further by relaxing the secrecy constraint for intermediate results.
As such, our approach can be applied only in cases where the disclosure of intermediate results is not regarded as critical --- a quite frequent case in practical applications.
Moreover, we aim 
at optimizing the sharing scheme for fast computation in the online phase.

When it comes to analyzing traffic data across multiple networks, various anonymization techniques have been
proposed for obscuring sensitive local information (e.g.,~\cite{Slagell2006Flaim}). However,
these methods are generally not lossless and introduce a delicate privacy-utility tradeoff~\cite{Pang2006tcpmkpub}.
Moreover, the capability of anonymization to protect privacy has recently been called in question, both from a technical~\cite{burkhart2010anonymization} 
and a legal perspective~\cite{Ohm2010}.

\section{Conclusions}

The use of SMC techniques has recently been proposed to overcome the inhibiting privacy concerns associated with inter-domain sharing of network traffic data. 
However, the cost at which the cryptographic privacy guarantees of SMC are bought is tremendous.
Although the design and implementation of basic SMC primitives have recently been optimized, processing time for queries is still in the order of several minutes and involves significant communication overhead.

In this paper, we further boost the performance of privacy-preserving network monitoring by two means. Firstly, we identify that perfect secrecy of intermediate results is not required in many cases.
That is, we advocate the use of ``elementary" (as opposite to ``complete") secure multiparty computation (E-SMC) procedures for traffic monitoring. E-SMC supports only simple computations with {\em private input and public output}, i.e., they can not handle secret input nor secret (intermediate) output. 
Secondly, we separate the computation into an \emph{offline} and an \emph{online} phase. 
Our proposed scheme \ata is based on additive secret sharing and pre-generates  random secret shares during the offline phase with only constant storage overhead. In the online phase, \ata allows to process actual queries with zero communication overhead. This enables adoption of SMC techniques on massive scales, both in terms of input data volume and number of participants. In the second part, we introduce a number of high-level primitives supported by \ata that cover a wide range of use cases in network monitoring, including the private generation of histograms, set operations, and anonymous publishing. 

In future work, we will evaluate \ata on real network setups and study hybrid approaches combining \ata with SSS to provide scalability 
 and functional completeness. \\

\section*{Acknowledgments} 
This work was supported by the DEMONS project funded by  the EU 7th Framework Programme [G.A. no. 257315] (\url{http://fp7-demons.eu}).

\bibliographystyle{abbrv}
\bibliography{reduce2themax}

\end{document}